\documentclass[11pt, a4paper]{article}

\usepackage[oldstyle]{hep-font} %

\usepackage{fullpage}

\usepackage{hep-math-font} %
\usepackage{hep-math} %
\numberwithin{equation}{section}

\usepackage{hep-bibliography} %
\bibliography{bibliography}

\usepackage{hep-float} %
\graphics{figures}

\usepackage{hep-text} %
\usepackage{hep-reference} %

\usepackage{hep-acronym} %

\usepackage{hep-title} %

\usepackage[feynman,plot]{hep-graphic}
\tikzsetexternalprefix{tikz/}
\graphicpath{./graphics}
\pgfplotsset{
 table/search path={data},
 legend style={font=\ostyle\footnotesize},
}
\usetikzlibrary{patterns.meta}
\usepgfplotslibrary{fillbetween}

\acronym{HNL}{heavy neutral lepton}
\acronym{LNC}{lepton number conserving}
\acronym{LNV}{lepton number violating}
\acronym{SM}{Standard Model}
\acronym{GUT}{grand unified theory}[grand unified theories]
\acronym{QFT}{quantum field theoretical}
\shortacronym{QCD}{quantum chromo dynamics}
\acronym{QM}{quantum mechanical}
\acronym{EW}{electroweak}
\acronym{EWSB}{electroweak symmetry breaking}
\acronym{VEV}{vacuum expectation value}
\acronym{SPSS}{symmetry protected seesaw scenario}
\acronym{LHC}{large hadron collider}
\acronym[HL-LHC]{HLLHC}{high luminosity phase of the \LHC}
\shortacronym{FCC}*{future circular collider}
\shortacronym{ATLAS}{A Toroidal \LHC ApparatuS}
\acronym{HL}{high luminosity}
\acronym{DOF}{degree of freedom}[degrees of freedom]
\acronym{LNLS}{\emph{lepton number}-like symmetry}
\shortacronym{CMS}{compact muon solenoid}
\acronym{LO}{leading order}
\acronym{CL}{charged lepton}
\acronym{NP}{new physics}
\acronym{MC}{Monte Carlo}
\acronym{PMNS}{Pontecorvo–Maki–Nakagawa–Sakata}
\acronym{pSPSS}{phenomenological symmetry protected seesaw scenario}
\acronym[$N\mkern-1.5mu\widebar N$O]{NNO}{heavy neutrino-antineutrino oscillation}
\acronym{PDF}{probability density function}
\acronym{LLR}{logarithm of the likelihood ratio}
\shortacronym{UFO}{universal \software{FeynRules} output}
\acronym[H.c.]{HC}{Hermitian conjugate}
\acronym[$0\nu$\beta\beta]{NLDB}{neutrinoless double \beta}
\acronym{BM}*{benchmark model}
\acronym{BR}{branching ratio}
\acronym{TOF}{time of flight}
\acronym{FCT}{Fundação para a Ciência e a Tecnologia}
\shortacronym{MLM}{Michelangelo L.\ Mangano}

\renewcommand{\vec}{\mathbf}
\NewDocumentCommand{\expno}{om}{\IfNoValueF{#1}{#1\times}10^{#2}}
\newcommand{\expnumber}[2]{\expno[#1]{#2}}

\newcommand{\ispd}{\differential{i\slashed\partial}}

\crefname{enumi}{step}{steps}

\usepackage{dcolumn}
\newcolumntype{d}[1]{D{.}{.}{#1}}
\newcommand{\header}[1]{\multicolumn{1}{c}{#1}}

\newcommand{\cutlabel}[2]{\textlabel[\usecommand{\MakeUppercase ##1}]{#1}[\bfseries]{#2}}
\newcommand\Cpp{C\nolinebreak[4]\hspace{-.025em}\raisebox{.25ex}{\smaller[.5]{++}}}

\title{Beyond lepton number violation at the HL-LHC: Resolving heavy neutrino-antineutrino oscillations}

\author[a]{Stefan Antusch \email{stefan.antusch@unibas.ch}}
\author[b]{Jan Hajer \email{jan.hajer@tecnico.ulisboa.pt}}
\author[a]{Johannes Rosskopp \email{johannes.rosskopp@unibas.ch}}

\affiliation[a]{Departement Physik, Universität Basel, Klingelbergstrasse 82, CH-4056 Basel, Switzerland}
\affiliation[b]{Centro de Física Teórica de Partículas (CFTP), Instituto Superior Técnico (IST), Universidade de Lisboa, Av.\ Rovisco Pais 1, 1049-001 Lisboa, Portugal}

\begin{document}

\maketitle

\begin{abstract}
Collider testable low-scale seesaw models predict pseudo-Dirac heavy neutrinos, that can produce an oscillating pattern of \LNClong and \LNVlong events.
We explore if such \NNOslong can be resolved at the $\HLLHC$.
To that end, we employ the first ever full \MClong simulation of the oscillations, for several example benchmark points, and show under which conditions the \CMS experiment is able to discover them.
The workflow builds on a \software{FeynRules} model file for the \emph{\pSPSSlong} (pSPSS) and a patched version of \software{MadGraph}, able to simulate \NNOslong.
We use the fast detector simulation \software{Delphes} and present a statistical analysis capable of inferring the significance of oscillations in the simulated data.
Our results demonstrate that, for heavy neutrino mass splittings smaller than about \unit[100]{\mu eV}, the discovery prospects for \NNOslong at the $\HLLHC$ are promising.
\end{abstract}

\clearpage
\tableofcontents
\listoftables
\listoffigures
\clearpage

\section{Introduction}

The observed light neutrino flavour oscillations \cite{SNO:2002tuh} can be explained by introducing at least two mass splittings between the light neutrinos.
It follows that at least two of the neutrinos have to be massive.
However, the \SM of particle physics lacks the corresponding mass terms.
Therefore, it needs to be extended with a model able to generate these light neutrino masses, such as the type~I seesaw mechanism \cite{Minkowski:1977sc, Gell-Mann:1979vob, Mohapatra:1979ia, Yanagida:1980xy, Schechter:1980gr, Schechter:1981cv}.
In order to generate the two light neutrino masses in this model at least two sterile neutrinos have to be added to the particle content of the \SM.
In general, the type~I seesaw mechanism is governed by two sets of parameters, that can be adjusted to obtain light neutrino masses at the right order of magnitude.
Besides their Yukawa couplings, the second set is given by their Majorana masses~$m_M^{}$, which are only possible since the additional neutrinos are \SM singlets.
After \EWSB the Yukawa couplings result in Dirac mass terms~$\vec m_D^{(i)}$ between each of the sterile neutrinos $N_i$ and the active neutrinos $\nu_\alpha$.
\footnote{We indicate quantities with a suppressed vector index by bold font.}
For the case of two heavy neutrinos, the generated light neutrino mass matrix then follows from the seesaw equation
\begin{equation} \label{eq:seesaw}
M_\nu = \frac{\vec m_D^{(1)} \otimes \vec m_D^{(1)}}{m_M^{(1)}} + \frac{\vec m_D^{(2)} \otimes \vec m_D^{(2)}}{m_M^{(2)}} \,.
\end{equation}
In order to obtain light neutrino masses, one could take the Yukawa couplings to be very small or choose the Majorana masses of the heavy neutrinos to be very large.
Those two limits of the seesaw mechanism are referred to as the small coupling and the high scale limit, respectively.
In both cases the parameters have to be taken to such extreme values that a direct detection of the sterile neutrinos at current collider experiments is not feasible.
A third possibility, which leads to collider testable low-scale seesaw models, consists in heavy neutrino pairs forming a pseudo-Dirac structure.
This leads to a cancellation between the two terms in \eqref{eq:seesaw} and can be justified by an approximate symmetry.
The symmetry can be realised as an extension of the \SM lepton number and we refer to it as a \LNLS \cite{Antusch:2022ceb,Antusch:2015mia,Antusch:2016ejd}.
It allows for the heavy neutrinos to be light enough, and at the same time have sufficiently large Yukawa couplings, to be directly observable at \eg the \LHC \cite{ATLAS:2015gtp, CMS:2018iaf, Drewes:2019fou, ATLAS:2020xyo, CMS:2022fut}.

When the \LNLS is exact, the light neutrinos are massless and the two heavy neutrinos are precisely mass-degenerate, forming a single Dirac neutrino.
However, if the \LNLS is broken by small parameters, light neutrino masses are generated and typically a mass splitting between the heavy neutrinos is introduced.
The corresponding seesaw scenario is referred to as \emph{\SPSS} \cite{Antusch:2022ceb, Antusch:2015mia, Antusch:2016ejd}.
The phenomenological effects of the slightly broken \LNLS are captured by the \pSPSS introduced in \cite{Antusch:2022ceb}.
Similar to the case of other neutral particles, this leads to particle-antiparticle oscillations \cite{Lande:1956pf, ARGUS:1987xtv, CDF:2006imy, LHCb:2012zll}, which are in this context called \NNOs \cite{Antusch:2017ebe}.
The \QFT framework of external wave packets, which has been developed in \cite{Beuthe:2001rc, Sachs:1963xxx,Kayser:1981ye, Giunti:1993se} and applied to the case of \NNOs in \cite{Antusch:2020pnn}, allows to describe this phenomenon including decoherence effects \cite{Antusch:2023nqd}.

Although the small breaking of the \LNLS generates \LNV processes, one can argue on principle grounds that for prompt heavy neutrino decays it is not possible to observe \LNV at the \LHC \cite{Kersten:2007vk}.
In contrast, \NNOs generate observable amounts of \LNV when the heavy neutrinos are sufficiently long-lived compared to their oscillation period.
When the heavy neutrinos have large enough lifetimes, such that their time of decay can be reconstructed, it can be possible to resolve the oscillation pattern of the \NNOs.
This allows to deduce the mass splitting of the heavy neutrinos, which might not be possible from measuring the amount of \LNC and \LNV processes alone.
The possibility of resolving the \NNOs at colliders has been estimated to be feasible during the \HLLHC in \cite{Antusch:2017ebe}.

The goal of the present paper is to explore under which conditions it is possible to resolve \NNOs at the \HLLHC at the reconstructed level, employing the process shown in \cref{fig:feynman}.
To this end, the \software{FeynRules} implementation of the \pSPSS \cite{FR:pSPSS} is used to simulate \NNOs in \software{MadGraph}.
Subsequently, \software{Pythia} is employed to simulate \QCD effects such as hadronisation, and the fast detector simulation \software{Delphes} is used to simulate the \CMS phase~II detector.
A cut based analysis of the generated events is performed using custom \software{\Cpp} code.
Finally, the prospects to resolve oscillation patterns for several \BM\ scenarios of long-lived nearly mass degenerate heavy neutrinos at the \HLLHC are derived, using a detailed statistical analysis implemented in \software{Mathematica}.

This paper is structured as follows:
First, the \pSPSS and \NNOs are briefly reviewed in \cref{sec:SPSS}.
Subsequently, in \cref{sec:simulation} the simulation of signal events containing these oscillations and relevant background processes are discussed.
Afterwards, the statistical analysis is introduced in detail in \cref{sec:statistical analysis}.
Finally, the results are presented in \cref{sec:results} and the paper is concluded in \cref{sec:conclusion}.
In \cref{sec:residual oscillations} we comment on the induction of residual oscillations through cuts on the transverse impact parameter.

\section{\sentence\SPSSlong} \label{sec:SPSS}

When considering two heavy neutrinos under the assumption of an intact \LNLS, the only allowed additions to the \SM Lagrangian are
\begin{equation} \label{eq:symmetric Lagrangian}
\mathcal L_{\SPSS}^L =
\widebar{N_i^c} \ispd N_i^{}
- y_{1\alpha} \widebar{N_1^c} \widetilde H^\dagger \ell_\alpha
- \widebar{N_1^c} m_M^{} N_2^{}
+ \dots + \HC \,,
\end{equation}
where $N_1$ and $N_2$ are sterile neutrinos written here as left-chiral four-component spinor fields.
The Higgs and lepton doublets of the \SM are denoted by $H$ and $\ell$, respectively, and $\vec y_1$ is the Yukawa coupling vector with components $y_{1\alpha}$.
The ellipses capture further contributions that can be generated by additional sterile neutrinos, but are assumed to be subdominant here.
In this case the two terms in the seesaw formula \eqref{eq:seesaw} cancel precisely.
In addition, the following symmetry breaking terms can be introduced
\begin{equation} \label{eq:broken Lagrangian}
\mathcal L_{\SPSS}^{\cancel L} =
- y_{2\alpha} \widebar{N_2^c} \widetilde H^\dagger \ell_\alpha
- \mu_M^\prime \widebar{N_1^c} N_1^{}
- \mu_M^{} \widebar{N_2^c} N_2^{}
+ \dots + \HC \,.
\end{equation}
When the Yukawa coupling vector $\vec y_2$ times the \VEV as well as the Majorana mass parameters $\mu_M^{}$ and $\mu_M^\prime$ are \emph{small} compared to $m_M^{}$, the light neutrino masses are guaranteed to be small as well.
Due to the approximate \LNLS, it is possible for the heavy neutrinos to have a mass well below the $W$ boson mass, while at the same time their coupling can be large enough to obtain a sizable number of events at \eg the \HLLHC, without violating constraints from light neutrino experiments such as searches for \NLDB decay.

\subsection{\sentence\pSPSSlong}

In the \SPSS with exact \LNLS \eqref{eq:symmetric Lagrangian}, the neutrino mass matrix for the interaction eigenstates $n = \row{\nu_e, \nu_\mu, \nu_\tau, N_1, N_2}^\trans$ is given by
\begin{equation}
M_n^L =
\begin{pmatrix}
0 & \vec m_D^{} & 0 \\
\vec m_D^\trans & 0 & m_M^{} \\
0 & m_M^{} & 0
\end{pmatrix}
\end{equation}
where $\vec m_D = \vec y_1 v$ with the \SM Higgs \VEV $v\approx\unit[174]{GeV}$.
The mass term of the Lagrangian after \EWSB is given by
\begin{equation}
\mathcal L_\text{mass} = - \frac12 \widebar{n^c} M_n n + \HC \,.
\end{equation}
This mass matrix does not generate neutrino masses.
Furthermore, the two heavy neutrinos are mass degenerate with a relative phase of $-i$ in their Yukawa couplings and can therefore be described as a single Dirac fermion.

However, in the presence of the small symmetry breaking terms \eqref{eq:broken Lagrangian} the complete mass matrix reads
\begin{equation} \label{eq:broken mass matrix}
 M_n^{\cancel L} =
\begin{pmatrix}
 0 & \vec m_D^{} & \vec \mu_D^{} \\
 \vec m_D^\trans & \mu_M^\prime & m_M^{} \\
 \vec \mu_D^\trans & m_M^{} & \mu_M^{}
\end{pmatrix}
\,,
\end{equation}
where $\vec \mu_D^{} = \vec y_2 v$.
This mass matrix not only generates small masses for the light neutrinos, but additionally introduces a mass splitting between the two heavy neutrinos, that is also suppressed by the same small \LNLS breaking parameters.
Thus, the pair of two Majorana fermions can no longer be described as a pure Dirac particle, but as a pseudo-Dirac particle.

The symmetry breaking terms can arise from specific low scale seesaw models, such as the linear or the inverse seesaw, which yield only the terms proportional to $\vec \mu_D^{}$ and $\mu_M^{}$, respectively.
However, the complete mass matrix \eqref{eq:broken mass matrix} can be generated from more complicated seesaw models.
Since the symmetry breaking terms are very small, their phenomenological effects can often be neglected in collider studies with the notable exception of \NNOs, which can be phenomenologically significant as they are an interference effect.

At \LO, the \NNOs are fully described by the mass splitting $\Delta m$ between the heavy neutrinos together with an additional damping parameter $\lambda$, capturing the potential decoherence effects discussed in \cref{sec:oscillations}.
Therefore, instead of adding the terms \eqref{eq:broken Lagrangian} to the Lagrangian \eqref{eq:symmetric Lagrangian}, in the \pSPSS, the mass splitting is directly introduced as a model parameter.
Consequently, the masses of the heavy neutrinos are given by
\begin{equation} \label{eq:Delta m mass splitting}
m_{\nicefrac45}^{} = m_M^{} \left(1 + \frac12 \abs{\vec \theta}^2 \right) \mp \frac12 \Delta m \,,
\end{equation}
where $\vec \theta = \flatfrac{\vec m_D^{}}{m_M^{}}$ is the active-sterile mixing parameter.
A detailed description of the \pSPSS can be found in \cite{Antusch:2022ceb}.

\subsection{\sentence\BMslong} \label{sec:benchmark}

\begin{table}
\begin{tabular}{cd{3.1}d{2.2}l} \toprule
\BM & \header{$\Delta m / \unit{\mu eV}$} & \header{$c \tau_\text{osc} / \unit{mm}$} & \header{$R_{ll}$} \\ \midrule
\ostyle 1 & 82.7 & 15 & 0.9729 \\
\ostyle 2 & 207 & 6 & 0.9956 \\
\ostyle 3 & 743 & 1.67 & 0.9997 \\
\bottomrule \end{tabular}
\caption[\sentence\BMlong\ points]{
All \BM\ points have a mass of \unit[14]{GeV} and an active-sterile mixing parameter satisfying $\abs{\theta_\mu}^2 = 10^{-7}$, which results for all points in a decay width of $\Gamma = \unit[13.8]{\mu eV}$.
However, they vary in their mass splitting and consequently have different oscillation periods $\tau_\text{osc}$, which leads to different \LNV to \LNC ratios $R_{ll}$, defined in \eqref{eq:Rll}.
} \label{tab:benchmark}
\end{table}

\begin{figure}
\includetikz{expected_events}
\caption[Number of expected events at the $\HLLHC$]{
Contour lines for the number of expected displaced vertex events $N$ as well as bands for the \LNV over \LNC event ratio $R_{ll}$, \cf definition \eqref{eq:Rll}, for the three \BMs\ defined in \cref{tab:benchmark}.
The contour lines for $N$ apply to the \CMS detector at the $\HLLHC$ with $\mathscr L = \unit[3]{\inv{ab}}$ and cuts as defined in \cref{sec:signal,sec:background}.
They have a nose-like shape and depend only on the heavy neutrino mass $m$ and the active-sterile mixing parameter $\theta_\mu$, with $\theta_e = \theta_\tau = 0$.
$N$ is therefore identical for all three \BMs, which differ only in their mass splittings $\Delta m$.
The position of the \BMs\ in the parameter plane is indicated by a purple cross.
The different $\Delta m$ of the \BMs\ result in different $R_{ll}$ bands.
The contour lines where $R_{ll} \in [0.1,0.9]$ are shown in yellow and orange for \BM1 and \BM3, respectively.
The relative position of the cross to the bands shows that the \BMs\ have an $R_{ll}$ close to one.
The grey area corresponds to the exclusion bounds from displaced vertex searches \cite{ATLAS:2020xyo, CMS:2022fut} and the shaded grey area to the bounds from searches for prompt \LNV processes \cite{ATLAS:2015gtp, CMS:2018iaf}.
Since the prompt searches rely on \LNV signals, they apply only to models with an $R_{ll}$ close to one.
} \label{fig:event number}
\end{figure}

In order to obtain the significance with which \NNOs could be observed at colliders, three \BMs\ that differ only in their mass splitting are used and given in \cref{tab:benchmark}.
The Majorana mass parameter and the active-sterile mixing parameters for all \BM\ points are chosen to be $m_M^{} = \unit[14]{GeV}$ and $\abs{\theta_\mu}^2 = \expno{-7}$, leading to a decay width of $\Gamma = \unit[13.8]{\mu eV}$.
The active-sterile mixing parameter corresponds to a Yukawa coupling of $y_\mu = \expnumber{2.55}{-5}$, while the Yukawa couplings to the electron and $\tau$-lepton have been set to zero.
The number of expected events with muons as final state leptons in conjunction with the cuts presented in \cref{sec:signal,sec:background}, are shown in \cref{fig:event number}.
It becomes clear that the chosen \BMs\ lie comfortably beyond the current bounds \cite{ATLAS:2020xyo, CMS:2022fut}.

In the minimal linear seesaw model, only one pseudo-Dirac pair of heavy neutrinos is added to the \SM.
This results in the lightest neutrino being massless and the mass splitting of the heavy neutrinos $\Delta m$ being identical to the mass splitting of the light neutrinos, \cf \cite{Antusch:2017ebe}.
The mass splitting of \BM3 is chosen to represent this possibility, where the light neutrino mass splitting is taken from a recent global fit assuming inverse light neutrino mass hierarchy \cite{Esteban:2020cvm}.
However, in top-down realisations of low scale seesaw models, \eg the low scale linear seesaw models in \cite{Malinsky:2005bi,Antusch:2017tud}, it is common to have multiple pseudo-Dirac pairs of heavy neutrinos.
Since the light neutrino masses get contributions from all these heavy neutrinos, it is possible for the pseudo-Dirac pairs to have smaller mass splittings than in \BM3 without the need of cancellations in the mass matrix.
For simplicity it might be assumed that the collider phenomenology is dominated by only one of the pseudo-Dirac pairs.
We introduce two additional \BM\ points reflecting this possibility.

\subsection{Oscillations} \label{sec:oscillations}

Oscillations of neutral particles, and thus \NNOs, can be described in the \QFT framework of external wave packets.
Compared to simple methods relying on plane waves, wave packets allow for a self-consistent description of oscillations.
This is due to the finite uncertainty in energy-momentum and spacetime, that allows to produce a coherent superposition of non-degenerate mass eigenstates, while simultaneously makes it possible to introduce the notion of a travelled time and distance.
The \QFT framework also allows to calculate the possible suppression of oscillations due to the loss of coherence between the propagating mass eigenstates.
For phenomenological studies these effects can be captured by a damping parameter $\lambda$, which is thus included in the \pSPSS.

At \LO, the probability to obtain a \LNC or \LNV event is given by
\begin{equation}
P^{\nicefrac{\LNC}{\LNV}}_\text{osc}(\tau) = \frac{1 \pm \cos(\Delta m \tau) \exp(-\lambda)}2 \,.
\end{equation}
Therefore, the oscillation period, defined in the proper time frame, is
\footnote{We use \emph{time} for quantities in the proper time frame and \emph{length} for quantities in the lab frame, independent from the units of those quantities.}
\begin{equation} \label{eq:oscillation period}
\tau_\text{osc} = \frac{2 \pi}{\Delta m} \,.
\end{equation}
Since the probability density of the heavy neutrino to decay is
\begin{equation}
P_\text{decay}(\tau) = - \dv \tau \exp\left(- \Gamma \tau\right) = \Gamma \exp\left(- \Gamma \tau\right) \,,
\end{equation}
the probability for a heavy neutrino to decay between the proper times $\tau_{\min}$ and $\tau_{\max}$ and forming an \LNC or \LNV event is given by
\begin{equation} \label{eq:probability displacement}
P_{ll}^{\nicefrac{\LNC}{\LNV}}(\tau_{\min}, \tau_{\max}) = \int_{\tau_{\min}}^{\tau_{\max}} P^{\nicefrac{\LNC}{\LNV}}_\text{osc}(\tau) P_\text{decay}(\tau) \d \tau \,.
\end{equation}
By integrating the oscillations from the origin to infinity, one derives the total ratio between \LNV and \LNC events \cite{Anamiati:2016uxp,Das:2017hmg,Antusch:2022ceb}
\begin{equation} \label{eq:Rll}
R_{ll} = \frac{P_{ll}^{\LNV}}{P_{ll}^{\LNC}} = \frac{\Delta m^2}{\Delta m^2 + 2 \Gamma^2} \,,
\end{equation}
that can be directly deduced from cut and count based analyses.
The expected number of events in a collider experiment is given by
\begin{equation} \label{eq:event numbers}
N^{\nicefrac{\LNC}{\LNV}} = \mathscr L \sigma \BR \int D(\vartheta, \gamma) P^{\nicefrac{\LNC}{\LNV}}_{ll}(\tau_{\min}(\vartheta, \gamma), \tau_{\max}(\vartheta, \gamma)) \d\vartheta \d\gamma\,,
\end{equation}
where the collider luminosity~$\mathscr L$, the heavy neutrino production cross section $\sigma$, and the \BR are used and $D(\vartheta, \gamma)$ is the probability density of the heavy neutrino to have a Lorentz factor $\gamma$ and an angle $\vartheta$ with respect to the beam axis.
The detector geometry is incorporated in the parameters $\tau_{\min}$ and $\tau_{\max}$ when boosting to the laboratory frame via $\tau(\vartheta,\gamma) = (\gamma^2-1)^{-\nicefrac12} L(\vartheta)$.
The \NNOs in the \pSPSS have been discussed in detail in \cite{Antusch:2022ceb}.
We have checked that for the parameter points under consideration decoherence effects can be neglected and we thus take $\lambda = 0$ in the following \cite{Antusch:2023nqd}.

\section{Simulation} \label{sec:simulation}

\begin{figure}
\includetikz*{oscillation-feynman-jet}
\caption[Diagram of heavy neutrino production, oscillation, and decay]{
Diagram depicting the production, oscillation, and decay of a heavy neutrino.
The heavy neutrino interaction eigenstate $N$ is produced in association with a prompt antimuon.
Subsequently, the mass eigenstates oscillate such that finally a neutrino or antineutrino interaction eigenstate decays into a displaced muon or antimuon, respectively.
We indicate that the process is initiated by proton collisions and that, for our parameter points, the two final quarks, originating from the hadronic $W$ decay, immediately hadronise into a single jet.
} \label{fig:feynman}
\end{figure}

In order to simulate events exhibiting \NNOs, we employ the \software{FeynRules} \cite{Alloul:2013bka} implementation of the \pSPSS \cite{FR:pSPSS}.
\software{FeynRules} is used to generate an \UFO output, passed to \software[2.9.10 (LTS)]{MadGraph5\_aMC@NLO} \cite{Alwall:2014hca} in order to generate events at parton level.
We use the patch introduced in \cite{Antusch:2022ceb}, that modifies the function calculating the particle's \TOF, to implement the oscillations.
When evaluating the matrix element, \software{MadGraph} treats the process as prompt, adding the \TOF information afterwards.
Therefore, almost no \LNV is obtained when interference between diagrams with different mass eigenstates are taken into account.
In order to circumvent this, the process is initially specified in such a way as to prevent interference between the heavy neutrino mass eigenstates.
This is achieved by writing the heavy neutrinos explicitly as intermediate states.
As a consequence the total cross sections for the \LNC and \LNV process are identical and the amount of generated \LNC and \LNV events is, except for statistical variations, the same.
The correct interference effects, including \NNOs, are then included via the patch described in \cite{Antusch:2022ceb}.
The syntax to generate a process, as the one given in \cref{fig:feynman}, is
\begin{verbatim}
generate p p > ww, (ww > mu nn, (nn > mu j j))
\end{verbatim}
where, in addition to the jet \code{j} and proton \code{p} multi particles, containing quarks and gluons, the multi particles
\begin{verbatim}
define mu = mu+ mu-
define ww = w+ w-
define nn = N4 N5
\end{verbatim}
are used.
Here, the initial $W$ boson is additionally taken to be on-shell, since we focus on heavy neutrinos with masses far below the $W$ boson mass.
An additional hard initial jet is included by using
\begin{verbatim}
add process p p > ww j, (ww > nn mu, (nn > mu j j))
\end{verbatim}
where \MLM matching was enabled with the standard parameter choices of \software{MadGraph} to prevent double counting.

\software{MadGraph} utilises \software[8.306]{Pythia} in order to hadronise and shower the parton level events \cite{Bierlich:2022pfr}.
Finally \software[3.5.0]{Delphes} is used with the standard \CMS phase~II card \nolinkurl{CMS_PhaseII_0PU} to simulate the detector effects \cite{deFavereau:2013fsa}.

\paragraph{Secondary vertex reconstruction and smearing}

Since \software{Delphes} neither simulates displaced tracks properly, nor reconstructs secondary vertices, we implemented a smearing function affecting the displaced vertices, with the idea to capture experimental uncertainties.
To our knowledge, no results for the overall precision of the vertex reconstruction have been published by \CMS so far.
Therefore, we vary it in our simulations between zero and \unit[4]{mm}, which we assume to be a conservative parametrisation of our ignorance, relying on private communication with members on the experimental collaborations.
We like to emphasise that it would be highly welcome if the experimental collaborations could provide such information and, ideally, if such uncertainties could be implemented in \software{Delphes} directly.
In detail, it is assumed that each reconstructed vertex is distributed with a Gaussian, with a standard deviation of $\unit[n]{mm}$, around its actual value.
The true value of the displaced vertex is obtained from the displaced muon.
The results are presented for different values of $n$, demonstrating the effects of the uncertainty in the vertex reconstruction.

\subsection{Signal} \label{sec:signal}

In order to observe \LNV via \NNOs using the process given by the diagram presented in \cref{fig:feynman}, the two leptons have to be measured.
For the \BMs\ given in \cref{tab:benchmark} and indicated in \cref{fig:event number}, the final state quarks are soft and immediately hadronise into a multi pion state, which then forms a single jet that can be captured with a cone radius of $\Delta R = 0.4$.
Furthermore, the \BM\ points are chosen such that the heavy neutrinos are long-lived.
Hence, the signal contains one prompt muon, one displaced muon, and one displaced jet.

A cut is used, which requires the muon to have an impact parameter of $\abs{d_0} \leq \unit[100]{\mu m}$, to ensure \cutlabel{cut:one prompt muon}{exactly one prompt $\mu$}.
Furthermore, this muon is required to have a transverse momentum $p_T^{}$ greater than \unit[20]{GeV}, ensuring that the event can be triggered.
Events that do not contain \cutlabel{cut:one disp muon}{at least one displaced $\mu$} are excluded by a cut that requires at least one muon with an impact parameter of $\abs{d_0} \geq \unit[2]{mm}$.
Furthermore, the displaced muon candidate is only valid if it emerges from the same vertex as the displaced jet.
A jet is considered displaced if it contains at least two displaced tracks that originate from the same vertex within a radius of \unit[100]{\mu m}.
A track is considered displaced if it has an impact parameter of $\abs{d_0} \geq \unit[2]{mm}$.
Signal events are required to have \cutlabel{cut:one disp jet}{exactly one displaced $j$}.

In addition to those basic cuts that define the signal, we demand \cutlabel{cut:muon isolation}{{}$\mu$ isolation} by requiring that the displaced muon and displaced jet have a $\Delta R$ larger than $0.4$ in order to reject muons radiated from the jet.
Furthermore, restricting the reconstructed heavy neutrino mass to the \cutlabel{cut:N mass window}{{}$N$ mass window} of $\pm \unit[2]{GeV}$ around its theoretical value results in a cleaner signal.
Generally, displaced particles are only considered if their origin falls inside a cylinder with dimensions given by half the tracker size in each direction.
For the \CMS phase~II detector, this means that displaced tracks must have an origin with a transverse distance smaller than \unit[60]{cm}, and a longitudinal distance smaller than \unit[150]{cm}, from the primary vertex.
This requirement ensures that the produced particles propagate through a volume of the tracker that should be large enough to facilitate detection.
At the moment this is an optimistic assumption about the performance of the displaced track reconstruction in the \CMS tracker, however, we think that pushing this capability is a worthwhile goal for the collaboration (see also \cite{Drewes:2019fou} and references therein).
Although the reconstruction of \NNOs discussed here mostly relies on displaced vertices appearing in the inner tracker, it is also possible to reconstruct muon tracks and vertices using the muon chambers \cite{Bobrovskyi:2011vx, Bobrovskyi:2012dc, CMS:2015pca}.

\subsection{Background} \label{sec:background}

\begin{table}
\begin{tabular}{llrr} \toprule
 & & \header{Signal} & \header{Background} \\
 & & & \header{Heavy hadrons} \\ \midrule
\multirow{2}{*}{Produced events} & Physical & $2854$ & $\expno[8.882]{7}$ \\
 & Simulated & $\expno[5]{4}$ & $\expno[1.1]{7}$ \\\midrule
\multirow{9}{*}{Cuts} & \ref{cut:one prompt muon} & $-23196$ & $-\expno[6.50]{6}$ \\
 & \ref{cut:one disp jet} & $-21652$ & $-\expno[4.48]{6}$ \\
 & \ref{cut:one disp muon} & $-1396$ & $-17411$ \\
 & \ref{cut:muon isolation} & $-838$ & $-45$ \\
 & \ref{cut:vertex direction} & $0$ & $0$ \\
 & \ref{cut:no prompt e} & $0$ & $0$ \\
 & \ref{cut:W mass window} & $-111$ & $-3$ \\
 & \ref{cut:N mass window} & $-1211$ & $0$ \\ \midrule
\multirow{2}{*}{Remaining events} & Simulated & $1596$ & $0$ \\
 & Physical & $91$ & $0$ \\
\bottomrule \end{tabular}
\caption[Cut flow of signal and background events]{
Cut flow for the signal and the heavy hadron background events.
The physical events are given at an integrated luminosity of $\mathscr L = \unit[3]{ab^{-1}}$ with a signal cross section of $\sigma = \unit[951]{ab}$.
} \label{tab:cut flow}
\end{table}

The cuts introduced above define the basic search strategy for the signal, which is based on reconstructing a displaced vertex.
The main sources of background for such a process are given by heavy flavour \SM processes generating long-lived heavy hadrons, interaction of particles with dense detector material, and cosmic ray muons \cite{ATLAS:2020xyo, CMS:2022fut}.

The heavy \SM hadrons can travel macroscopic distances before they decay, potentially forming displaced vertices.
However, this is only possible if those heavy flavour particles, and consequently their decay products, are highly boosted.
Since the selection rules that define the signal already require the displaced muon and the displaced jet to have a $\abs{d_0} > \unit[2]{mm}$, this background is strongly reduced \cite{CMS:2014xnn, ATLAS:2020xyo}.
A \cutlabel{cut:W mass window}{{}$W$ mass window} cut around the reconstructed $W$ boson mass of $ \pm \unit[20]{GeV}$ is employed to reduce the background even further.
We simulated this background using the same programs as for the signal events.
The \software{MadGraph} syntax we used is
\begin{verbatim}
generate p p > mu all bb
\end{verbatim}
where the additional multi particle
\begin{verbatim}
define bb = b b~
\end{verbatim}
is defined.
An additional hard initial jet was included by using
\begin{verbatim}
add process p p > mu all bb j
\end{verbatim}
where \MLM matching was enabled with the standard parameter choices of \software{MadGraph}.
The whole process was simulated using the model \code{sm-no_b_mass}, which employs the five flavour scheme in the definitions of the protons and jets.
After the parton level events are passed to \software{Pythia} and \software{Delphes}, the above mentioned cuts and event selection rules were used, and it has been found that the background is eliminated entirely.
The cut flow for the simulated signal and background is given in \cref{tab:cut flow}.

Interactions of \SM particles with detector material can also result in a displaced vertex signature and thus have to be considered as part of the background.
A map of regions containing dense detector material is required in order to accept only displaced vertices that are outside that region.
Simulating this background requires detailed knowledge about the detector structure and is beyond the scope of this work.
However, requiring a displaced vertex to be reconstructed from enough \emph{good} tracks reduces this background.
In an experimental analysis, a sophisticated track reconstruction algorithm judges which tracks are good, in the sense that the track is likely to be produced from a charged particle rather than detector noise.
With no specific insight about this algorithm, we define the tracks as being good that pass our cut and selection rules, which ensure that the charged particles traverse at least half the tracker.
With the cuts introduced to define the signal, it is ensured that each event contains at least three displaced tracks from which the displaced vertex can be reconstructed.

In order to veto against cosmic ray muons, the \cutlabel{cut:vertex direction}{vertex direction} cut is implemented that requires the reconstructed heavy neutrino momentum to be in the same direction as the displaced vertex.
For this the distance in $\row{\eta, \phi}$-space between the reconstructed momentum $\vec p_N^{}$ and the displaced vertex direction $\vec d$ is limited to $\Delta R \leq 1.5$.
We assume that this background is eliminated when using this cut in combination with additional timing information, therefore we have not attempted to simulate it.

With the cuts described above, the signal region can be assumed to be background free.
Additionally we found that a \cutlabel{cut:no prompt e}{prompt $e$ veto}, where prompt is again defined as $\abs{d_0} < \unit[100]{\mu m}$, with a $p_T^{} \geq \unit[20]{GeV}$ does not affect the signal, such that it could be used to eliminate further background if that becomes necessary.

\section{Statistical Analysis} \label{sec:statistical analysis}

\newcommand{\lh}{P}
\newcommand{\lhr}{L}
\newcommand{\llhr}{\Lambda}
\newcommand{\pdf}{\mathcal P}

The number of events entering the cut based analysis is given by the luminosity times the cross section for the considered process.
The number of expected events after the cut based analysis $N_\text{exp}$ is then given by
\begin{align} \label{eq:number of events}
N_\text{exp} &= \mathscr L \sigma f_\text{eff} \,, &
f_\text{eff} &= \frac{N_\text{after cuts}}{N_\text{all events}}\,,
\end{align}
where the efficiency factor captures how the cuts summarised in \cref{tab:cut flow} impact the number of signal events.
The set of events surviving all cuts is labeled $\mathcal D_\text{all}$, and can be divided into \LNC events $\mathcal D_{\LNC}$ and \LNV events $\mathcal D_{\LNV}$.
These datasets are then binned in the proper time $\tau$ space and the number of events in the $i$-th bin are given by
\begin{equation}\label{eq:bin counts}
N_i = \abs*{\set*{E \in \mathcal D \suchthat \tau_E^{} \text{ in $i$-th bin}}}\,,
\end{equation}
where $\abs{}$ is the cardinality such that \eg $N_\text{exp} = \abs{\mathcal D_\text{all}}$.
The heavy neutrino \TOF $\tau_E^{}$ is defined as the proper time at which it decays in the event $E$.

\subsection{Hypotheses}

The shape of the histograms describing the heavy neutrino \TOF may be predicted by two hypotheses.
In contrast to the null hypothesis, the alternative hypothesis features oscillations.

\paragraph{Null hypothesis}

\newcommand{\plotref}[1]{(\ref*{#1})} %
\begin{figure}
\includetikz[.618]{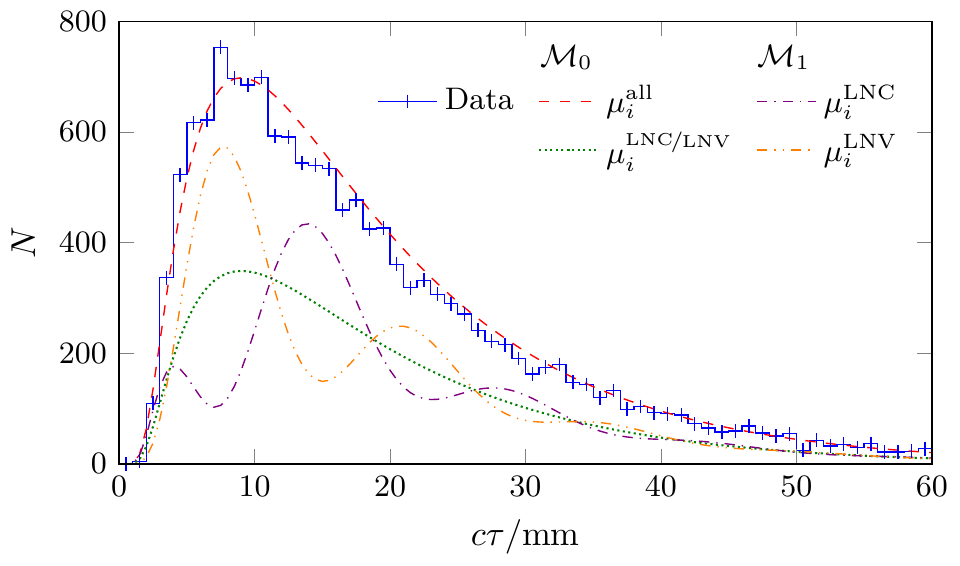}
\caption[Examples for the null and alternative hypotheses]{
Example histogram \plotref{plot:M0} with \unit[15]{k} events, demonstrating the proper time distribution after applying the cuts summarised in \cref{tab:cut flow}.
The inverse Gaussian distribution $\mu_i^\text{all}(\mathcal M_0)$ described in \eqref{eq:inverse Gaussian} is depicted \plotref{plot:M0 fit} together with the prediction of the null hypothesis $\mu_i^{\nicefrac{\LNC}{\LNV}}(\mathcal M_0)$ \plotref{plot:M0 fit half}.
The inverse Gaussian is shown with the parameters given in \cref{it:one} on \cpageref{it:one}, where the normalisation is approximated as $N_0 = 15500$.
The distribution of the alternative hypothesis \eqref{eq:model 1} is shown for \LNC \plotref{plot:M1 LNC} and \LNV \plotref{plot:M1 LNV} events of \BM1, with example value $\alpha = 0.05$ for the washout parameter.
} \label{fig:hypotheses}
\end{figure}

The hypothesis without oscillations $\mathcal M_0$ is based on the assumption that, for each bin, the probability of the heavy neutrino superposition to decay in a \LNC process is equal to the probability to decay in a \LNV process.
Therefore, the mean number of predicted events in the $i$-th bin is given by
\begin{equation}
\mu_i^{\nicefrac{\LNC}{\LNV}}(\mathcal M_0) = \frac12 \mu_i^\text{all}(\mathcal M_0) \,.
\end{equation}
The probability of a given bin count can then be computed assuming a Poisson distribution of bin counts around this mean.
In principle, one would expect the predicted mean values of this hypothesis to follow an exponential due to the finite lifetime of the heavy neutrinos.
However, the employed cuts change this distribution into a non trivial one, which can be approximated by the \PDF of a generalized inverse Gaussian, described by four free parameters, yielding
\newcommand{\param}[1]{\mathsf#1}
\begin{equation} \label{eq:inverse Gaussian}
\mu_i^\text{all}(\mathcal M_0) = \frac{N_0}{2} \frac{\tau_i^{\param\theta - 1}}{\param\mu^{\param\theta} K_{\param\theta}(\nicefrac{\param\lambda}{\param\mu})} \exp\left(-\frac{\param\lambda}{\param\mu^2} \frac{\tau_i^2 + \param\mu^2}{2 \tau_i}\right)\,,
\end{equation}
where $N_0$ denotes the overall normalisation, $\param\theta$ is the index parameter, $\param\mu$~is the mean of the distribution, and $\param\lambda$ is a scale parameter.
Additionally, $\tau_i$ denotes the position of the middle point of the $i$-th bin, and $K_{\alpha}(z)$ is the modified Bessel function of the second kind.
The distribution is shown in \cref{fig:hypotheses}, where the parameters correspond to the best fit point estimated in \cref{it:one} on \cpageref{it:one}.

\paragraph{Alternative hypothesis}

The second hypothesis $\mathcal M_1$ includes oscillations, with an oscillation period given by \eqref{eq:oscillation period}, on top of the distribution described by the first hypothesis.
Due to the imperfect reconstruction of the Lorentz factor, an additional washout effect obscures the oscillation pattern for larger~$\tau$.
This effect is included by an exponential factor $\alpha$, suppressing the oscillation.
The prediction for the mean number of expected events in a bin is then given by
\begin{equation} \label{eq:model 1}
\mu_i^{\nicefrac{\LNC}{\LNV}}(\mathcal M_1) = \left[1 \pm \cos(\Delta m \tau_i) e^{- \alpha \tau_i} \right] \mu_i^{\nicefrac{\LNC}{\LNV}}(\mathcal M_0) \,,
\end{equation}
where the two additional parameters $\Delta m$ and $\alpha$ are incorporated, resulting in a total of six free parameters.
Again, a Poisson distribution of bin counts, around this mean, is assumed.
One example of such oscillations is given in \cref{fig:hypotheses}.

\subsection{Likelihood ratio test}

To test whether the hypothesis including oscillations is statistically preferred by the simulated data, we use a likelihood ratio test.
The main idea is to decide if the null hypothesis, given by $\mathcal M_0$, can be discarded in favour of the alternative hypothesis, given by $\mathcal M_1$.
The likelihood is denoted by $\lh(\set{N_i}, \mathcal M)$ and describes the probability of finding the measured bin counts $\set{N_i}$ given the hypothesis $\mathcal M$.
It is given by the product of likelihoods for all bins in both, the \LNC and \LNV, cases.
With the assumed Poisson distributed number of events around the mean value $\mu_i$ in the $i$-th bin, the likelihood of a single bin is given by
\begin{equation}
\lh(N_i, \mu_i) = \frac{{\mu_i}^{N_i} e^{-\mu_i}}{N_i !} \,,
\end{equation}
where, as introduced above, $N_i$ is the number of events in the $i$-th bin, which depends on the dataset and the binning.

The best fit point for each hypothesis, given the bin counts $\set{N_i}$, is evaluated by maximising the corresponding likelihood.
A hypothesis at its best fit point is denoted with a hat, \eg $\widehat{\mathcal M}_0$.
Therefore, the likelihood ratio can be computed, given the bin counts, and yields
\begin{equation}
\lhr(\set{N_i})
= \frac{\lh(\set{N_i}, \widehat{\mathcal M}_0)}{\lh(\set{N_i}, \widehat{\mathcal M}_1)}
= \prod_i \frac{\lh(N_i^{\LNC}, \widehat{\mathcal M}_0) \lh(N_i^{\LNV}, \widehat{\mathcal M}_0)}{ \lh(N_i^{\LNC}, \widehat{\mathcal M}_1) \lh(N_i^{\LNV}, \widehat{\mathcal M}_1)} \,.
\end{equation}
Taking the washout parameter to infinity in the alternative hypothesis reproduces the null hypothesis.
The two hypotheses are therefore nested and as a consequence the inequality
\footnote{
This inequality can by violated if not a global but a local maximum of the likelihood is found, while searching for $\widehat{\mathcal M}_1$.
Such a local maximum might result in a smaller likelihood for the alternative hypothesis than the likelihood obtained by a fit of the null hypothesis.
}
\begin{equation}
\lh(\set{N_i}, \widehat{\mathcal M}_0) \leq \lh(\set{N_i}, \widehat{\mathcal M}_1) \,,
\end{equation}
holds and hence the likelihood ratio is restricted by
\begin{equation}
0 \leq \lhr(\set{N_i}) \leq 1 \,.
\end{equation}
A likelihood ratio close to zero means that the given binned data are much better fitted by the alternative hypothesis than by the null hypothesis.
In contrast, a ratio close to one shows that there is no clear distinction between the two hypotheses for the given binned data.
Since in practice the logarithm of the likelihood is better suited for numerical computations, we continue the discussion using the \LLR which is defined as
\begin{equation}
\llhr(\set{N_i}) = -2 \log(\lhr(\set{N_i})) \,.
\end{equation}
The \LLR ranges from zero to infinity, where now a value near zero states that both, the null hypothesis and the alternative hypothesis, produce an equally good fit of the binned data.
Contrary, a high value states that the alternative hypothesis produces a better fit.

\paragraph{Probability}

\begin{figure}
\begin{panels}{3}
\smaller[.5]
$\llhr=5.70$, $p=\unit[21.5]{\%}$, $Z=\unit[0.79]{\sigma}$
\includetikz{oscillations-low}
\caption{Low significance}\label{fig:low}
\panel
\smaller[.5]
$\llhr=13.59$, $p=\unit[0.82]{\%}$, $Z=\unit[2.4]{\sigma}$
\includetikz{oscillations-mid}
\caption{
Intermediate significance}\label{fig:mid}
\panel
\smaller[.5]
$\llhr=17.71$, $p=\unit[0.13]{\%}$, $Z=\unit[3.01]{\sigma}$
\includetikz{oscillations-high}
\caption{High significance}\label{fig:high}
\end{panels}
\caption[Statistical fluctuation of the null hypothesis]{
Three examples of statistical fluctuations in the null hypothesis, producing patterns that mimic \NNO.
From panel \subref{fig:low} to panel \subref{fig:high} the oscillatory pattern becomes more distinct while the probability that the depicted histogram is generated decreases.
} \label{fig:null hypothesis}
\end{figure}

Caused by statistical fluctuations of the bin counts around their predicted mean, the null hypothesis $\widehat{\mathcal M}_0$ can also feature an oscillation pattern, as demonstrated in \cref{fig:null hypothesis}.
Therefore, the value of a \LLR alone does not contain enough information to decide whether oscillations in a given dataset are significant or not.
To translate the \LLR into a significance, it is crucial to know the probability that the null hypothesis produces the same \LLR due to fluctuations.
Per construction, the most likely \LLRs produced by $\widehat{\mathcal M}_0$ are small, and the larger the \LLR the less likely it is to be produced.
The distribution of \LLRs, generated by statistical fluctuations, is the \PDF of the \LLR under the assumption that the null hypothesis is true.
We label it $\pdf$.

\begin{table}
\begin{tabular}{ccc} \toprule
\BM & $\chi^2$ $\DOFs$ & $k(\unit[5]{\sigma})$ \\ \midrule
\ostyle 1 & 2.25 & 30.94 \\
\ostyle 2 & 3.28 & 34.05 \\
\ostyle 3 & 3.92 & 35.82 \\
\bottomrule \end{tabular}
\caption[Null hypothesis fluctuations]{
Statistical fluctuations in the bin counts, following the null hypothesis, lead to a $\chi^2$ distribution with given $\DOFs$.
The distributions are computed following the \cref{it:one,it:two,it:three,it:four} on \cpageref{it:one}.
The \LLR corresponding to a $p$ value of \unit[5]{\sigma} are found to be increasing from \BM1 to \BM3.
} \label{tab:chi2 distributions}
\end{table}

The challenge is then to find a value $k_p$, for which the probability of obtaining a $\llhr(\set{N_i}) \geq k_p$ through statistical fluctuations, is smaller than $p$.
This means that $p$ is the probability of rejecting the null hypothesis with respect to the alternative hypothesis, even though the null hypothesis is true.
Given $\pdf$, the value of $k_p$ can be obtained via
\begin{equation}
\int_{k_p}^{\infty} \pdf(k) \d k = p \,.
\end{equation}
While in the limit of large sample sizes the \PDF is typically assumed to approach a $\chi^2$ distribution, the form of the \PDF for finite sample sizes is generally unknown.
Therefore a simulation is performed, with the goal to sample the \PDF, as follows
\begin{enumerate}
\item\label{it:one} All simulated events that survive the cut based analysis are used to estimate the true distribution of the null hypothesis $\widehat{\mathcal M}_0$, which is depicted in \cref{fig:hypotheses}.
For those events, the best fit parameters of $\widehat{\mathcal M}_0$ are given by $\param\mu = 10.81$, $\param\lambda = 17.28$, and $\param\theta = 0.71$.
The overall normalisation parameter $N_0$ is not relevant for the following steps as they only depend on the \PDF, which is normalised to unity.

\item\label{it:two}
To obtain a set of events that follows the null hypothesis, taking statistical fluctuations into account, \TOFs are taken according to the \PDF from \cref{it:one}.
This is done until the set contains $N_\text{exp}$ signal events.

\item\label{it:three} Using these \TOF values, bin counts are computed using the binning parameters based on \cref{eq:bin counts}.

\item\label{it:four} The \LLR is computed based on the bin counts from \cref{it:three}.
\end{enumerate}
By repeating \cref{it:two,it:three,it:four} the desired distribution can be obtained.

The obtained distributions are well approximated by $\chi^2$ distributions, where the \DOFs are treated as a free parameter.
The values of the \DOFs for the three \BMs\ are given in \cref{tab:chi2 distributions}.

\paragraph{Significance}

\begin{figure}
\begin{panels}[t]{2}
\includetikz{oscillations-BM1}
\caption{
The simulated events are based on \BM1 with an oscillation period of \unit[15]{mm}.
The best fit parameters are given by an oscillation period of $\unit[14.08^{+0.85}_{-0.71}]{mm}$ and a washout parameter of $\alpha = \expnumber{3.66^{+31.73}_{-3.66}}{-3}$.
The $\LLR$ was found to be $51.0$, which in this case yields a significance of \unit[6.66]{\sigma}.
}
\panel
\includetikz{oscillations-BM3}
\caption{
The simulated events are based on \BM3 with an oscillation period of \unit[1.67]{mm}.
The best fit parameters are given by an oscillation period of $\unit[1.63^{+0.03}_{-0.04}]{mm}$ and a washout parameter of $\alpha = \expnumber{7.44^{+17.76}_{-5.45}}{-2}$.
The $\LLR$ was found to be $5.29$, which in this case yields a significance of \unit[0.67]{\sigma}.
}
\end{panels}
\caption[Examples for the statistical fit of the oscillations to the $\MC$ data]{
Examples for the best fit of the alternative hypothesis to the data.
For each parameter point a luminosity of \unit[3]{\inv{ab}} is used, resulting in a total of about $90$ events after cuts.
However, the number of events contributing to the fit is based on the range of $\tau$ values used as given in \cref{tab:binning} and therefore differs between the \BMs.
The fit has been performed based on the binning options in \cref{tab:binning}.
In these examples the secondary vertex smearing has been neglected.
The bands around the oscillations depict the errors of one standard deviation, assuming a Poisson distribution for the event count in each bin.
} \label{fig:oscillations}
\end{figure}

With the so obtained \PDF it is possible to compute the probability that a given \LLR produced from statistical fluctuations of the null hypothesis $\widehat{\mathcal M}_0$.
This probability $p$ can be translated to a significance $Z$ by comparing it to a standard normal distribution.
A smaller probability translates to a larger significance and a higher threshold $k_p$.
A significance of \unit[5]{\sigma} corresponds to a probability of $p \approx \expno[2.87]{-7}$, which for \BM1 corresponds to a \LLR threshold of $k_p \approx 30.94$.
See \cref{tab:chi2 distributions} for the values of the other \BMs.
Therefore, in the case of \BM1, a \LLR greater than $30.94$ can be interpreted as a discovery, where oscillations have been found with a significance larger than \unit[5]{\sigma}.
For very small \LLRs, it is possible that the obtained probability is greater than \unit[50]{\%}.
Since for the translation into a significance a standard Gaussian is used, \cf \cite{ParticleDataGroup:2022pth}, the corresponding significance would be smaller than zero .
However, in cases where the \LLRs are so small, the result should just be interpreted in the way that no oscillations could be proven in the given data.
The simulated events and the resulting statistical fit for two example oscillations are shown in \cref{fig:oscillations}.

\subsection{Data pre-processing}

\begin{table}
\begin{tabular}{cccc} \toprule
\BM & $c \tau_\text{max}/\unit{mm}$ & Bins & Used events \\ \midrule
\ostyle 1 & 60 & 30 & 87 \\
\ostyle 2 & 60 & 30 & 85 \\
\ostyle 3 & 15 & 60 & 42 \\
\bottomrule \end{tabular}
\caption[Binning parameters]{
Binning parameters for the different \BM\ points, as well as the number of events used in the computation of the $\LLR$.
Note that for small samples of event surviving the cut based analysis, as it is the case here, the number of events participating in the computation of the $\LLR$ undergoes fluctuations, which contribute to the fluctuations in the obtained significances.
} \label{tab:binning}
\end{table}

Some pre-processing of the events is performed to stabilise the numerics.
In principle, it is expected that for sufficiently large proper times alternative hypotheses with a washout parameter greater zero produce the same bin counts as the corresponding null hypotheses.
Therefore, the \LLRs are dominated by small proper times, for which the washout effect is also small.
We have found that the fitting algorithm used by us gives more reliable results when we consider only these dominant oscillations.
Therefore we do not use all events for the statistical analysis but consider only a window containing the first oscillations.
The details of this restriction for each \BM\ point are given in \cref{tab:binning}.
To help reduce noise in the data a Gaussian filter of radius $r = 1$, which corresponds to a standard deviation of $\sigma = \flatfrac{r}{2}$ in bin space, is applied to the bin counts before the maximisation takes place.

\section{Results} \label{sec:results}

In this paper, the first ever analysis of \NNOs at reconstructed level is performed.
A detailed statistical analysis is employed to obtain a significance describing the feasibility to resolve the oscillations at the \HLLHC.
To that end, three \BM\ points with different oscillation periods, given in \cref{tab:benchmark}, have been simulated.
All \BM\ points feature a mean mass of the heavy neutrinos of \unit[14]{GeV} as well as an active-sterile mixing parameter of $\abs{\theta_\mu}^2 = 10^{-7}$.
This leads to a decay width of $\Gamma = \unit[13.8]{\mu eV}$.
The parameters are chosen such that the bounds of current collider searches are well evaded.
While \BM3 captures the mass splitting of the minimal linear seesaw model, producing the measured light neutrino data with an inverted hierarchy, \BM1 and \BM2 feature a smaller mass splitting as discussed in \cref{sec:benchmark}.

\paragraph{Secondary vertex smearing}

\begin{figure}
\begin{panels}[t]{2}
\includetikz{significance-smearing}
\caption{Secondary vertex smearing for $90$ signal events.} \label{fig:smearing}
\panel
\includetikz{significance-luminosity}
\caption{Lorentz factor reconstruction error in \BM3.} \label{fig:reconstruction error}
\end{panels}
\caption[Significance as function of vertex smearing and Lorentz factor reconstruction]{
Panel \subref{fig:smearing}:
Significance of the three \BM\ points at a luminosity of $\mathscr L = \unit[3]{ab^{-1}}$ as a function of the secondary vertex smearing.
Panel \subref{fig:reconstruction error}:
Significance of \BM3, as function of the number of events surviving the analysis, for three different relative errors of the reconstruction of the Lorentz factor \eqref{eq:gamma error}.
For this comparison no smearing has been taken into account.
} \label{fig:scan variation}
\end{figure}

Since \software{Delphes} does not provide the experimental uncertainty to reconstruct secondary vertices, we have introduced a free parameter governing Gaussian smearing of the otherwise perfectly reconstructed secondary vertices.
\Cref{fig:smearing} shows the expected confidence with which oscillations could be observed in the data for a given smearing of the secondary vertex.
With our \BM\ points, the number of signal events that survive the cuts, shown in \cref{tab:cut flow}, is roughly $90$.
This assumes a luminosity of $\unit[3]{ab^{-1}}$ which is the expected total integrated luminosity of the \HLLHC \cite{ZurbanoFernandez:2020cco}.
A factor of two in the number of events can easily be achieved by choosing \BM\ points closer to the excluded region.
Therefore we regard this analysis as conservative.
For each data point in the plot, $100$ \LLRs have been computed to obtain a mean value and a standard deviation for the significance.
The figure shows that for an oscillation period of \unit[15]{mm} in proper time space, corresponding to \BM1, a significance of \unit[5.19]{\sigma} can be expected if no smearing is taken into account.
Moreover, the significance is above \unit[5]{\sigma} up to a smearing of \unit[2]{mm}, dropping to \unit[4.46]{\sigma} for a smearing of \unit[4]{mm}.
Parameter points with smaller oscillation periods are affected stronger by the smearing.
This is expected since a smaller oscillation period in proper time space is related to a smaller oscillation length in lab space.
A larger smearing therefore results in a stronger washout for smaller oscillation periods.

\paragraph{Lorentz factor reconstruction}

Since it is crucial to reconstruct the Lorentz factor on an event by event basis, we show in \cref{fig:reconstruction error} how a better reconstruction of the Lorentz factor as well as a higher luminosity makes it possible to observe oscillations even for \BM3.
The reconstruction error directly effects the quality of the reconstructed oscillations in the proper time frame.
Therefore, if the error is too large for the mass splitting one wants to resolve oscillations for, a higher event numbers only yields a limited improvement.
This is shown by the lowest line in \cref{fig:reconstruction error}.
In contrast the higher lines, that represent a smaller reconstruction error of the Lorentz factor, benefit much more from a larger event number.

The quality of the reconstruction of the Lorentz factor is measured using the relative error of the Lorentz factor, which is defined as
\begin{equation} \label{eq:gamma error}
\Delta \gamma = \frac{\abs{\gamma_\text{gen}^2 - \gamma_\text{reco}^2}}{\gamma_\text{gen}^2 - 1} \,,
\end{equation}
where $\gamma_\text{gen}$ is the true Lorentz factor of the heavy neutrino and $\gamma_\text{reco}$ is the reconstructed one.
The set of events used for the analysis yields an exponential distribution of the relative errors of the Lorentz factor.
Most events have a small $\Delta \gamma$, while only a small fraction of events have a large $\Delta \gamma$.
The overall quality in the reconstruction can be measured by the standard deviation of that exponential distribution.
While a large standard deviation corresponds to many events with a large relative error, and therefore to a poor reconstruction, the opposite is true for a small standard deviation.
Without any improvement, with respect to the Lorentz factor achieved in this analysis, the events of \BM3 yield $\Delta \gamma$'s that result in an exponential distribution with a standard deviation of $0.16$.
By improving the reconstruction such that the standard deviation is reduced to $0.0096$, the significance is improved from around zero to $\unit[(3.37 \pm 1.10)]{\sigma}$ for $90$ events.
Additionally doubling the number of events yields a significance of $\unit[(5.13 \pm 2.28)]{\sigma}$.
This scaling is justified since we assume that it is possible to improve the \LHC analysis presented here, such that more signal events survive while the background is still eliminated.
Additionally, it is possible to choose a \BM\ point closer to the experimentally excluded region as mentioned earlier.
Furthermore, future collider experiments, such as the ones at the \FCC\ \cite{FCC:2018evy, FCC:2018vvp}, might yield much higher luminosities as well as better reconstruction possibilities of the Lorentz factor of the heavy neutrino.

\paragraph{Mass splitting dependency}

Oscillations can be used to resolve very small mass splittings.
However, from the discussion above it is clear that if the mass splitting becomes too large, \ie the oscillation period becomes too small, the reconstruction of oscillation patterns will be challenging.
This is shown in \cref{fig:oscillation period} where the oscillation period has been varied, using a fast simulation described below.
One can see that larger oscillation periods produce higher significances.
It is expected that the significance drops again if the oscillation period reaches the mean lifetime of the heavy neutrinos, since then oscillations can not develop before the mass eigenstates decay.

The fast simulation is based on the assumption that the kinematics of the events is independent of the oscillation period.
With this assumption, the oscillation period has no impact on the cut based analysis.
As a consequence, the sum of \LNC and \LNV events follows for all \BM\ points the same distribution, given by the null hypothesis.
Thus it is possible to obtain a large sample of valid signal events by combining the events passing all cuts of the three simulated \BM\ points.
It is then possible to give each event a new \emph{tag}, describing if that events should be counted as \LNC or as \LNV.
For this the \TOF of the heavy neutrino is computed on a per event basis, using the relation
\begin{equation}
\tau = \frac{\abs{\vec d}}{\sqrt{\gamma^2 - 1}}\,,
\end{equation}
where $\vec d$ is the position of the displaced vertex with respect to the primary vertex.
The formula for the oscillation probability can then be used to tag the event based on the new oscillation period.
At this point one has generated a sample of valid signal events with the new oscillation period.
After that, one can pick a random subset of this sample containing the physical number of events, that can be computed using \eqref{eq:number of events}.
Subsequently, the analysis to obtain the significance can be applied to this subset of events.
This strategy is orders of magnitude faster than performing the full \MC simulation and cut based analysis for each oscillation period separately.

\section{Conclusion} \label{sec:conclusion}

\begin{figure}
\includetikz[.618]{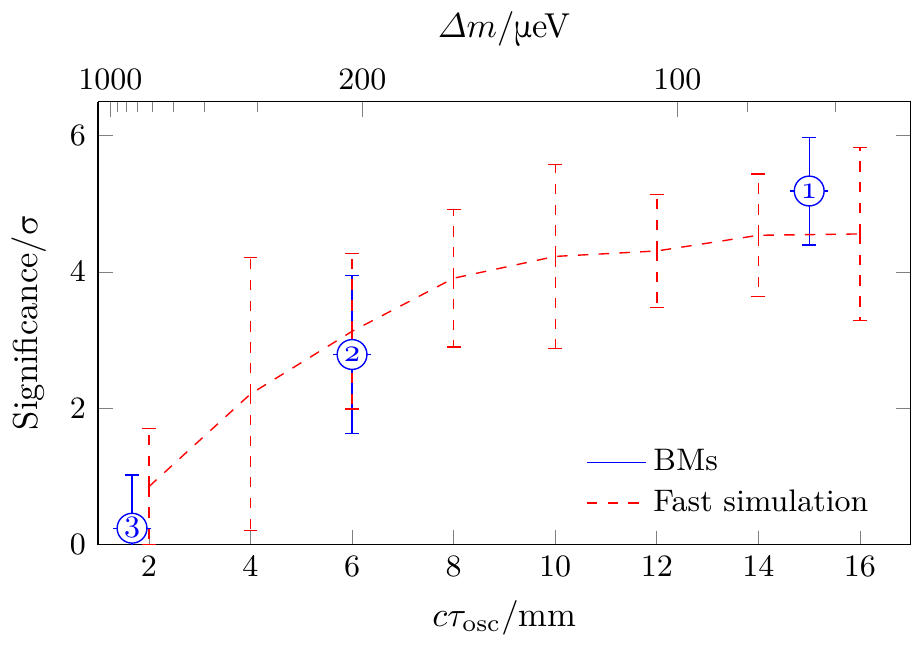}
\caption[Significance as function of the oscillation period]{
Significance as function of the oscillation period for $90$ events.
Three independently simulated \BM\ points as well as eight points simulated using a fast simulation are depicted.
Smearing of the secondary vertices is neglected.
} \label{fig:oscillation period}
\end{figure}

In this paper, we have performed a first full analysis of \NNOs at the reconstructed level.
The simulations are based on the \software{FeynRules} implementation of the \pSPSS introduced in \cite{Antusch:2022ceb}.
After the generation of events at parton level with a patched version of \software{MadGraph}, hadronisation and showers are simulated using \software{Pythia}.
Subsequently, a fast detector simulation of the \CMS detector has been preformed using \software{Delphes}.
The uncertainty in the reconstruction of the displaced vertex has been implemented with a smearing function, that randomly selects a value of the displaced vertex around its true value, based on a Gaussian.
The analysis of the events is performed using custom \software{\Cpp} and \software{Mathematica} code.

In our analysis we have focused on three \BMs\ within the \pSPSS, defined in \cref{tab:benchmark}, with heavy neutrino parameters conservatively chosen inside the region allowed by current experimental constraints.
The \BMs\ differ by the heavy neutrino mass splitting, which is largest for \BM1 and smallest for \BM3.
Simulating events containing heavy \SM hadrons, we have shown that with the event selection rules and cuts as defined in \cref{sec:signal,sec:background} the corresponding background is completely evaded.
It has also been argued that with the given cuts other backgrounds that could not be simulated should also be evaded.
Thus, the surviving signal events can be treated as background-free.
The statistical method to obtain the significance with which oscillations can be found in the simulated data is described in \cref{sec:statistical analysis}.

Our analysis shows that for small enough heavy neutrino mass splittings, corresponding to large enough oscillation periods, it is possible to discover \NNOs with the \CMS detector at the \HLLHC assuming \unit[3]{\inv{ab}} integrated luminosity.
The impact of the oscillation period, the displaced vertex smearing, the number of events, and the error in the reconstruction of the heavy neutrino Lorentz factor on the significance are shown in \cref{fig:scan variation,fig:oscillation period}.
For resolving the \NNOs, it is important that the smearing is smaller than the oscillation length in lab space.
Similarly, the variance of the \TOF, due to the error in reconstructing the Lorentz factor, should be smaller than the oscillation period.
This is the case for \BM1, for which a significance of $\unit[(5.01 \pm 0.9)]{\sigma}$ is obtained, assuming a smearing of \unit[2]{mm} and around $90$ total events relevant for the analysis, \cf \cref{tab:binning}.
For smaller oscillation periods, as in \BM2 and \BM3, the significance is below \unit[3]{\sigma} even if smearing is not taken into account.

However, we like to stress that smaller mass splittings might also be resolved with higher significance if the reconstruction of the Lorentz factor is improved.
This would not only increase the significance itself but also improve the effect for larger event numbers as shown in \cref{fig:reconstruction error}.
The event number could, \eg, be increased by choosing a parameter point closer to the experimentally excluded region, \cf \cref{fig:event number}.
Additionally, it might be possible to increase the significance for smaller mass splittings by increasing the decay width of the heavy neutrinos, \ie parameter points with increased Yukawa couplings or mass.
Then the heavy neutrinos would decay faster and there would be more events in the first oscillation cycles such that resolving the pattern becomes more feasible.
However, more events would be lost by the $d_0$ cut in this case.
In order to study the interdependence between such considerations, a scan over a larger sample of benchmark parameters is necessary.
In addition, the presented study might be improved by more sophisticated background reduction and by optimising the window of considered \TOFs defined in \cref{tab:binning}.

In summary, we have shown that the \HLLHC offers the exciting possibility to not only discover the \LNV induced by \NNOs of long-lived heavy neutrinos, but also to resolve the \NNO pattern.
Reconstructing the oscillation period would allow to measure this mass splitting and therefore discover the pseudo-Dirac nature of the heavy neutrino pair.
This would provide deep insight into the mechanism of neutrino mass generation and could help to shed light on whether leptogenesis is able to generate the baryon asymmetry of the universe (as discussed \eg in \cite{Antusch:2017pkq}).

\subsection*{Acknowledgements}

The work of J.H.\ as partially supported by the Portuguese \FCT through the projects CFTP-\FCT Unit UIDB/\allowbreak00777/\allowbreak2020, UIDP/\allowbreak00777/\allowbreak2020, CERN/\allowbreak FIS-PAR/\allowbreak0002/\allowbreak2021, and CERN/\allowbreak FIS-PAR/\allowbreak0019/\allowbreak2021, which are partially funded through POCTI (FEDER), COMPETE, QREN and the EU.
S.A.\ and J.R.\ acknowledge partial support from the Swiss National Science Foundation grant 200020/175502.

\appendix

\section{Residual oscillations} \label{sec:residual oscillations}

\begin{figure}
\begin{panels}[t]{2}
\includetikz{residual}
\caption{Impact of a $d_T$ cut}
\label{fig:dT}
\panel
\includetikz{residual-2}
\caption{\LNV oscillation pattern after a $d_0$ cut}
\label{fig:d0}
\end{panels}
\caption[Residual oscillations induced by a $d_0$ cut]{
Effects of the spin correlation sensitive cut $d_0$, in comparison to the non-sensitive cut $d_T$, on a set of $N_{\LNC} + N_{\LNV}$ events.
For each of the three datasets, represented by histograms, $\expno[5]{5}$ generator level events, based on the values of \BM3, have been simulated.
All histograms are normalised with the same factor, that ensures that the area under the uncut histogram in \subref{fig:dT} sums to unity.
Panel \subref{fig:dT}: Comparison of the \MC data before and after a $d_T$ cut of \unit[4]{mm}.
Panel \subref{fig:d0}: Impact of a $d_0$ cut of \unit[4]{mm} overlayed by the pattern of \LNV oscillations.
Due to the angular dependence appearing in \eqref{eq:impact parameter} the $d_0$ cut does not only generate a delayed onset but additionally imprints an oscillatory pattern originating in the \LNV oscillations.
} \label{fig:residual oscillations}
\end{figure}

The transverse impact parameter $d_0$, calculated for displaced tracks, is not only proportional to the transverse lifetime of the decaying particle, but additionally contains a component depending on the angle between the displaced vertex direction and the observed particle momentum.
If the magnetic field can be neglected, which we explicitly checked to be the case for the \BM\ points discussed in this paper, the transverse impact parameter is given by \cite{Antusch:2022ceb}
\begin{equation} \label{eq:impact parameter}
d_0 = d_T \sin(\varphi(\vec p_T^N, \vec p_T^\mu))\,,
\end{equation}
where $d_T$ is the transverse distance of the displaced vertex and the sine measures the angle in the transverse plane between the momenta of the heavy neutrino and the muon it decays into.
This sine introduces an angular dependency, sensitive to spin correlations in the process under consideration.
Since the \LNC and \LNV processes expose dissimilar spin correlations, the $d_0$ cut effects them differently.
This leads to the observation of \NNOs patterns in event samples that are a priory insensitive to the difference between \LNC and \LNV processes.
As an example, \cref{fig:residual oscillations} shows the residual oscillations appearing in a large sample of $N_{\LNC} + N_{\LNV}$ events after introducing a $d_0$ cut.
While the event sample with no cuts does not feature any oscillation pattern, it is shown that the $d_0$ cut results in residual oscillations, with peaks aligning with the ones of the \LNV oscillation pattern.
It can be concluded that the $d_0$ cut affect the \LNC events more severely than the \LNV ones.
By contrast, a cut on $d_T$ is independent of spin correlations and thus no residual oscillations appear.
We found that this effect is subdominant for smaller event samples, such as in this analysis, and therefore neglected it in the main part of the paper.

\printbibliography

\end{document}